\documentclass[twocolumn,9pt]{extarticle} 

\usepackage[twoside=true, head=1pc,
  paperwidth=8.5in, paperheight=11in,
  includeheadfoot, columnsep=2pc,
  top=57pt, bottom=73pt, inner=54pt, outer=54pt
]{geometry}
\usepackage{amsthm, amsmath, amsfonts, hyperref, algorithmic, algorithm}
\usepackage{graphicx}
\usepackage{subcaption}

\usepackage{booktabs}

\DeclareMathOperator*{\argmax}{arg\,max}

\title{\Huge{A supervised approach to time scale detection in dynamic networks}}

\author{\Large{Benjamin Fish}\thanks{University
of Illinois at Chicago, Department of Mathematics, Statistics, and Computer
Science, bfish3@uic.edu}\and \Large{Rajmonda S. Caceres}\thanks{MIT Lincoln Laboratory, Lexington, Massachussetts}}

\begin{document}

%
\maketitle

\begin{abstract}
For any stream of time-stamped edges that form a dynamic network, an important choice is the aggregation granularity that an analyst uses to bin the data.  Picking such a windowing of the data is often done by hand, or left up to the technology that is collecting the data.  However, the choice can make a big difference in the properties of the dynamic network.  This is the time scale detection problem.  In previous work, this problem is often solved with a heuristic as an unsupervised task.  As an unsupervised problem, it is difficult to measure how well a given algorithm performs.  In addition, we show that the quality of the windowing is dependent on which task an analyst wants to perform on the network after windowing.  Therefore the time scale detection problem should not be handled independently from the rest of the analysis of the network.

We introduce a framework that tackles both of these issues:  By measuring the performance of the time scale detection algorithm based on how well a given task is accomplished on the resulting network, we are for the first time able to directly compare different time scale detection algorithms to each other.  Using this framework, we introduce time scale detection algorithms that take a supervised approach: they leverage ground truth on training data to find a good windowing of the test data.  We compare the supervised approach to previous approaches and several baselines on real data.
\end{abstract}
\section{Introduction}
Much big data mining on social and other types of networks either requires information about dynamics or is improved by such information.  Incorporating temporal information can improve the efficacy and lead to more detailed analyses.


As data collection becomes cheaper and easier, the rate at which the data is being collected is often orders of magnitude more frequent than the underlying system.  The rate of the data collection process is typically a function of the technology used and not necessarily related to the evolution or dynamics of the network itself.  Thus the data collection process makes a choice about the bin size - also called the \emph{resolution}, \emph{aggregation granularity}, or \emph{time scale} of the dynamic network - that may not be the correct choice.  Binning the data at courser resolutions may make it possible to distinguish between noisy local temporal orderings and critical temporal orderings in the network.    Indeed, the time scale of the network strongly impacts what structures and dynamics may be observed in the network~\cite{clauset2012persistence,krings2012effects,ribeiro2013quantifying}.  Moreover, the choice of time scale impacts the efficacy of data mining on networks~\cite{fish2015handling}.  Thus for any data mining task over dynamic networks, choosing the bin size is not only important, but a necessary choice the data scientist must make, and leaving it up to the data collection process is not ideal.  This is the problem we take up in this paper, variously called time scale detection, generating graph snapshots, oversampling correction, temporal scale inference, aggregation granularity detection, or windowing detection.  We will refer to it as \emph{windowing} to emphasize the fact that the bin sizes - or windows - may not necessarily all be the same size.

This problem is marginally related to change point detection, which is the problem of detecting when the network changes drastically, inducing a segmentation of the sequence, and is only a matter of degrees away from windowing detection, which also asks for a segmentation.  Implicitly, however, change point detection assumes the network is observed at the right time scale.  

Typically, the windowing problem is framed as an unsupervised task.  If the goal is data exploration and it is unclear how we will use the data in the future, then unsupervised windowing may be sufficient.  On the other end of the spectrum, if sufficient knowledge of the data is at hand, the time scale may be chosen manually to represent natural scales, such as the diurnal or weekly scales natural for dynamics among people.  However, frequently, getting such domain-knowledge through a data exploration phase or otherwise may be prohibitively difficult or expensive.  Moreover, we will demonstrate that the best windowing often cannot be defined independently of the analytical task.


In this paper, we show that the time scale for a dynamic network depends on the task - the most appropriate choice of scale for predicting new links appearing in the network may be different than the appropriate choice for detecting change points in the network, for example.  The intuition is that the information needed to predict new links is different than the information needed to detect change points.  This insight may appear quite intuitive, but it runs contrary to many approaches for problems typically framed as unsupervised.


This informs our approach towards windowing:  we set up the windowing problem as a supervised machine learning task.  For example, for detecting change points, we set aside earlier training data from the network with labeled change points.  We then find the right time scale for this labeled data, and apply the time scale to new data to find the change points there.  In other words, the windowing algorithm takes as a parameter the desired task, such as change point detection, and finds the right time scale for that task on the input data.  Of course, the downside to this approach is that it requires training data, such as labeled change points on past data.  However, we regard this as both a natural assumption and a necessary assumption:  Many popular prediction and classification problems have some notion of ground truth.  In addition, since the best time scale depends on the task anyway, we might as well tailor our time scale detection towards a particular goal.

In this paper, we consider link prediction, attribute prediction, and change point detection.  We use these three tasks as popular examples of tasks analysts perform on dynamic networks that have some notion of ground truth.

Our contributions are as follows:  In addition to giving evidence that the best windowing of the data is task-dependent, we, for the first time, describe a simple framework for directly comparing the performance of windowing algorithms that leverages task-dependency.  We also introduce windowing algorithms that takes a supervised-machine-learning approach.  We compare our approach against several baselines and previous work.  We demonstrate that this supervised approach is often superior to other approaches, but that like all supervised approaches in machine learning, its quality may be dependent on the quality of the training data.

\subsection{Previous Work}

In numeric time series, this problem is often termed `segmentation,' and segmentation of numeric times series  has a long history which is outside the scope of this work; see~\cite{keogh2004segmenting} for an overview.

For dynamic networks, there has been some work related to our problem in the area of change point detection, which seeks to find points in time where the dynamic network has changed abruptly.  Typical methods include using generative models of dynamic networks~\cite{peel2015detecting} or clustering similar time slices ~\cite{berlingerio2010time}.  This literature is marginally different from the problem addressed in this paper because - while finding change points do implicitly segment the dynamic network - the goal of change point detection is not necessarily to find a good representation of the network for the purpose of binning each segment but rather just to find the points when the dynamic network undergoes significant change.

Also related is graph compression, which tries to find a representation of the dynamic network that minimizes the bits needed to store it while simultaneously retaining sufficient information about the network, under the general heading of graph summarization algorithms.  See~\cite{liu2016graph} for a survey of graph summarization techniques.  This approach has been applied, more generally, to multi-layer networks, dynamic or not~\cite{de2015structural}.  This problem is also slightly different from ours because we do not necessarily seek a small representation, merely an accurate representation for the task at hand.

Meanwhile, previous work on our problem has focused either on task-independent heuristics or methods that attempt to optimize for a specific metric on graphs.  Caceres et al.~maps the dynamic network to a time series using a metric on graphs (such as the number of triangles in each graph) and then determines time scale by finding the scale at which the time series' compression ratio and variance is balanced~\cite{sulo2010meaningful}.  Soundarajan et al.~determines a windowing of the data with respect to some metric (such as the exponent of the degree distribution) by measuring when that metric has converged~\cite{soundarajan2016generating}.  In our view, these approaches have the downside that they require a specific metric.  For a given data set or task to accomplish on a data set, it is not necessarily clear what metric to choose.  Darst et al.~use a parameter-free approach that seeks to find an appropriate time scale by measuring the similarity between graphs using the Jaccard index~\cite{darst2016detection}.  Most closely related to the approach that we take in this paper, Fish and Caceres use the quality of the performance of link prediction algorithms to determine the best time scale~\cite{fish2015handling}.  These are parameter-free methods or, relatedly, methods that assume that there is some `ground-truth' time scale via a generative model or the like, as in~\cite{clauset2012persistence,darst2016detection,fish2015handling}.  In this paper, we do not take this tactic because as we demonstrate, choice of time scale may be dependent on the task at hand, which functions as a parameter for the problem.

\subsection{Background}

Consider a dynamic network over a fixed set of vertices $V$, represented as a stream of time-stamped edges.  The goal of windowing is to segment this input stream of edges into (possibly overlapping) intervals to form a sequence of graphs $H_1,\ldots,H_m$, each graph representing all of the edges that occurred within each interval.  Representing the input edge stream as a sequence of graphs $G_1$,\ldots,$G_T$, a \emph{window} of size $k$ is a sequence $G_i,G_{i+1},\ldots,G_{i+k-1}$.  A \emph{windowing} is a sequence of windows $\{G_1,\ldots,G_{k_1}\},\{G_{k_1}+1,G_{k_1}+2,\ldots,G_{k_2}\},\ldots,\{G_{k_{m-1}},\ldots,{G_T}\}$.  In general, it is possible to also consider windows that overlap, but in this paper we focus on non-overlapping windows.  The resulting sequence is $H_1,\ldots,H_m$, where $H_i = \cup_{j=k_{i-1}+1}^{k_i} G_j$.  It is possible to consider other functions mapping a window to the resulting graph $H_i$, but we only consider the simple union in this paper.  As a slight abuse of notation, we will refer to both the segmentation of the input graph sequence and the resulting sequence $H_1,\ldots,H_m$ as a windowing.  

If all windows have the same size $w$ (except possibly the last window if the length of the sequence does not divide $w$), we refer to this as a \emph{uniform windowing} and $w$ the \emph{bin size}, \emph{window size}, or \emph{time scale}.  A window size of $w=1$ represents the time scale of the collection process and a window size of $w=T$, where $T$ is the duration of the observed network,  means that all temporal information is ignored.  The goal of this paper is to describe and evaluate windowing algorithms, i.e.~algorithms for finding a windowing given an input sequence of graphs.  As pointed out in~\cite{fish2015handling,soundarajan2016generating,sulo2010meaningful}, too small a window size may introduce noise and lack the structure necessary for analysis but too large a window size may lose important temporal information.

Once we have found a windowing $H_1,\ldots,H_m$, it can now be the input for any task that operates over a dynamic network.  We consider three popular tasks:  link prediction, attribute prediction, and change point detection.

In this paper, we consider a supervised approach:  the best windowing is the windowing that maximizes the performance of the algorithm for a given task, which we will refer to as the \emph{task algorithm}.  This gives us a way to compare different windowing algorithms:  we evaluate the performance of the windowing algorithm by evaluating the performance of the task algorithm on a test set.  In general, we are given the edges of a dynamic network up to some time $t$ as a training set, and performance is evaluated on the dynamic network from time $t+1$ onwards.  The windowing algorithm gets ground truth on the training set, e.g.~the change points that occurred, and the task algorithm uses only the windowed graph sequence to conduct its analysis, say finding the change points in the test set.

\section{Tasks}\label{sec:tasks}
Given a candidate windowing, we evaluate its quality by performing a learning task algorithm on that windowing.  We then evaluate the windowing by the performance of the task algorithm when using that windowing.  We consider three task algorithms:  link prediction, attribute prediction, and change point detection.  We treat link prediction as an online task, and attribute prediction and change point detection as offline tasks.  We do this to demonstrate windowing algorithms on both kinds of tasks.  In what follows, we describe the algorithms we use for each task and how we judge their performances.

\subsection{Link prediction}
In link prediction, the goal is to predict the edges that are most likely to appear in the future.  In the online setting, at every time step, our goal is to predict the edges that will appear in the next time step (the next step in the initial input sequence before windowing).

While there are many methods for link prediction (see~\cite{al2011survey} for a survey), the method we use is a simple scoring function that scores every pair of vertices by how likely an edge is to appear between them.  In this paper, we use the $\text{Katz}_\beta$ score, an efficient and well-performing score~\cite{LibenNowell2003}.  $\beta$ is a damping parameter that weights shorter paths exponentially higher than longer paths in the most recent graph in the input sequence.  For our experiment results, we use $\beta=0.005$, which has been used before~\cite{fish2015handling,LibenNowell2003}.

The performance of the link prediction algorithm is evaluated using the AUC of the precision-recall curve, as recommended by~\cite{yang2015evaluating}, averaged over all predictions made, one set of predictions for each graph.  

\subsection{Attribute prediction}
We use the Time Varying Relational
Classifier (TVRC) algorithm of Sharan and Neville~\cite{sharan2008temporal} to determine the unknown value of a binary vertex attribute.  As in their work, we assume attributes do not change over time.  The goal is then to infer the missing attribute values  by taking advantage of not only the known attribute values of the vertices but also by using the temporal information in the data.  Their method is a Bayesian model that takes advantage of knowledge of attributes in a vertex's neighborhood.  This model uses the time stamps to weight the influence each vertex has on its neighbors.  In our implementation, we use add-one smoothing for categorical features and a Gaussian distribution for continuous features.  The goal is then to find a windowing where TVRC builds the best performing model.  TVRC requires a kernel for weighting the importance of edges - as they suggest, we use their exponential kernel $(1-\theta)^{t-i}\theta$, where $t$ is the total time, $i$ the current time, and $\theta$ a hyperparameter controlling the rate of decay (for the sake of simplicity, we set $\theta=1/2$ without also trying to optimize this parameter). 


We use a special form of leave-one-out testing to measure the performance of TVRC.  In this setting, the target attribute of one of the vertices is removed from both training and testing, a model is trained with the training set, and gives a prediction for the value of the missing attribute using the test set.  To make the problem harder and more realistic, instead of just removing one target attribute, we remove a whole batch of them at once, and use the trained model to predict the values of all of their target attributes.  The vertex set is partitioned into batches using a batch size parameter $b$, and this is repeated for each batch.  Once a prediction has been made for all batches, we measure the performance or TVRC as the standard AUC of the ROC curve.

\subsection{Change point detection}
We use Graphscope, from Sun et al.~\cite{sun2007graphscope}.  Graphscope detects change points by estimating the times where segmenting the graph sequence at those times maximizes compressibility.  Since our graphs are not bipartite, we make the necessary modifications to Graphscope so that it may be used in the non-bipartite case.

In order to evaluate a change point detection algorithm, we need a single score summing up how good a set of change points are, but to the best of our knowledge no such measure has been proposed in the literature, contrary to classification tasks, which frequently use AUC or other single values to summarize quality. We will need a single score rather than, say, a precision-recall curve, not only to evaluate, but also because our supervised windowing algorithm will need a single score to directly and automatically compare the quality of different window sizes.

To rectify this, we propose the following measure:  Let $t_1,\ldots,t_k$ be the times of the ground-truth events, and $s_1,\ldots,s_\ell$ the times of the proposed events.  Let $n$ be the length of the sequence and $\delta(x)$ the function that equals $1$ if $x$ is true and $0$ otherwise.  Peel and Clauset~\cite{peel2015detecting} propose the following notions of precision and recall:
\[\text{Precision}(d)=\frac{1}{\ell}\sum_{i} \delta(\inf_j |s_i - t_j| \le d)\] 
\[\text{Recall}(d)=\frac{1}{k}\sum_{j} \delta(\inf_i |s_i - t_j| \le d).\]

We will define the AUC of the precision-recall curve as the normalized volume under the curve given by $\text{Precision}(d)$ and $\text{Recall}(d)$ as $d$ goes from $0$ to $n$.  Consider the set of distances $\{|s_i-t_j|:i,j\} \cup\{0,n\}$, ordered from smallest to largest as $d_1 < \ldots < d_m$, so $d_1=0$ and $d_m=n$.  Then PR-AUC is defined as
\[\frac{1}{n}\sum_{i=1}^{m-1} (d_{i+1}-d_i)\cdot\text{Precision}(d_i)\cdot\text{Recall}(d_i).\]
For the sake of completeness, if $k=0$ or $\ell=0$, we define the PR-AUC to be $0$.
Note this is a $[0,1]$-valued measure.

\section{Experimental Setup}\label{sec:exper_setup}

We now describe how we test the performance of windowing algorithms.
In the offline setting, we set aside a previous interval of the dynamic network for training, and the next interval for testing.  The training interval includes ground-truth information.  For example for change point detection, it includes any change points that occurred in that interval.  A windowing algorithm may use this information to decide on a window size to use in the test set.  A windowing algorithm, uniform or otherwise, is also allowed to see the edges in the test set to determine the windowing for the test set, but of course no ground truth information about the task.  

Once the test set is windowed, we perform our task on the windowed data and measure its performance, as describe in Section~\ref{sec:tasks}.  The windowing algorithm's score is then just the score that the task algorithm received.  For each data set, we split it up into six consecutive intervals, and then do training and testing on consecutive intervals, where the previous test set becomes the next training set, so there are five total tests\footnote{For Reality Mining, on change point detection, we split it into five instead of six intervals, in order to have every training set contain at least one change point.}.  We do this in order to promote generalizability of our results.  For change point detection, we merely average the scores over each of the tests.  For attribute prediction, we use pooling:  we take the AUC as described in Section~\ref{sec:tasks} over all vertices in all test sets, instead of averaging the individual AUC's of each test set, because the population, i.e.~the vertices, is the same in each test set.  

In the online setting, a new graph from the initial input sequence is given to the windowing algorithm, and the windowing algorithm must make a decision as to how to incorporate the new graph into the windowing so far.  After windowing, a prediction is made, and then the process is repeated.  Since the link prediction algorithm we use only ranks the likelihood of an edge appearing rather than determining the number of new links, we only do this process for those time steps where at least one new edge appears.  The score for the windowing algorithm is the average over all scores received for each prediction.  As in the offline setting, we split each data set into six consecutive intervals and then do training and testing on each pair of intervals, where the testing phase is online.

\begin{table*}
\caption{The score in the $i$th row and $j$th column is the score for the $j$th task if the window size is chosen to maximize the score for the $i$th task.  The first three columns are for the Enron data set, the last three are for the Reality Mining data set.  Scores are averaged over the different intervals of the data.}
\begin{tabular}{ccccccccc}
\toprule
& & Enron & & & & Reality Mining &\\
\midrule
 & Link prediction & Attr. prediction & CP detection & & Link prediction & Attr. prediction & CP detection\\
 \midrule
Link prediction & \bf{0.188} & 0.599 & 0.572 &  & \bf{0.277} & 0.961 & 0.778\\
Attr. prediction & 0.163 &\bf{0.649} & 0.540 &  & 0.220 & \bf{0.983} & 0.340\\
CP detection & 0.141 & 0.609 & \bf{0.935} &  & 0.258 & 0.961 & \bf{0.945}\\
\bottomrule
\end{tabular}
\label{table:best}
\end{table*}

\begin{table*}
\caption{Spearman correlation coefficients and p-values for each pair of tasks.}
\centering
\begin{tabular}{cccc}
\toprule
 & Link prediction & Attribute prediction & CP detection\\
  & \& Attribute prediction & \& CP detection & \& Link prediction\\
 \midrule
Enron & (0.093, 0.005) & (0.233, 5.11e-13) & (-0.355, 4.19e-29)\\
Reality Mining & (-0.050, 0.493) & (-0.188, 0.010) & (-0.071, 0.330)\\
\bottomrule
\end{tabular}

\label{table:spearman}
\end{table*}

\section{Data sets}\label{sec:data}
We use five data sets:  Enron, MIT Reality Mining, Badge, Hypertext09, and Haggle.  We treat all of these as undirected dynamic networks.  For both convenience and uniformity, we bin each of these data sets at a initial window size at a `natural' size, i.e.~a choice that a data analyst might make, such as an hour or a day.  This is so that a window size of $w=1$ represents a baseline representing how well a hand-chosen windowing would perform.  We only consider window sizes at least as large as this baseline.

Each of these are suitable for link prediction.  Of these, Enron, Reality Mining, and Badge are equipped with vertex attributes in order to test attribute prediction, and of these, Enron and Reality Mining also have established change point information.

\subsection{Enron}
This is an email network between employees of Enron Inc. from January 1999 to July 2002, during the period of Enron's market manipulation scandal and subsequent collapse~\cite{klimt2004enron}.  We use the emails of 151 employees, where each edge is an email sent from one of those employees to another.  Each vertex, representing an employee, have a binary attribute indicating whether the employee was a manager or not, and fifty integer-valued attributes, which are the number of occurrences in each employee's outgoing emails of the top fifty words used in all emails (a list of stop words were excluded from the top fifty words).  The attribute we test on for attribute learning is whether the employee was a manager or not, which we took from~\cite{creamer2009segmentation}.  We use the change points from~\cite{peel2015detecting}, which represent times when the emails undergo substantial shifts, such as the launch of Enron online and changes in the CEO position.  The initial bin size is one day.

\subsection{Reality Mining}
This is a proximity network of 90 MIT students and faculty using data taken from cell phones from September 2004 to May 2005 (we only use data up until the end of the academic year)~\cite{eagle2006reality}.  We use as edges both phone calls between participants and whenever two participants are close to each other, detected using Bluetooth.  Each of the participants filled out a survey about their cell phone usage, such as how much they use their cell phone, where they live etc., which we use as the categorical attributes for each vertex.  The attribute we use for testing attribute prediction is whether they are part of the business school or the MIT Media Lab.  Change points are taken from~\cite{peel2015detecting}, which are the start and ends of vacations, semesters, etc.  The initial bin size is one day.

\subsection{Badge}
This is a proximity network of 23 employees at a data server configuration firm for a month (the name of the data set comes from the badges the employees wore to track their location at the workplace)~\cite{olguin2009badge}.  Each edge represents when two employees are in close proximity to each other, representing an interaction.  Each employee was assigned a certain number of tasks, and data about these tasks was recorded, e.g.~average completion time, whether they took on a difficult task or not, etc.  For attribute prediction, we predict whether or not they made an error in one of their tasks.  The initial bin size is one hour.

\subsection{Haggle Infocomm}
This is a proximity network consisting of interactions, recorded using Bluetooth, among attendees at an IEEE Infocomm conference over four days~\cite{haggle}.  41 attendees participated in this network.  The initial bin size is 10 minutes.

\subsection{Hypertext09}
This is another proximity network of attendees at the ACM Hypertext 2009 conference, held over three days~\cite{isella2011hypertext}.  Each vertex is one of the 113 attendees, and an edge represents a interaction between two attendees that was active for at least 20 seconds.  The initial bin size is 10 minutes.

\section{Task dependence}

Ideally, it would be nice to have just one windowing algorithm that performs well regardless of whether your goal is link prediction, attribute prediction, change point detection, or any other task.  However, we demonstrate that this does not appear to be feasible while still maximizing the performance of the task algorithm.

To do this, we score each window size by the performance of each of the three task algorithms when the data set is windowed at that size, so we have a score representing the quality of the window size for each of the three tasks.  Table~\ref{table:best} show the score for the $i$th task, if you chose the window size with the highest score for the $j$th task.  (Scores are computed for each of the intervals as described in Section~\ref{sec:exper_setup} and the values in Table~\ref{table:best} are the average of the results for each interval).  For example, choose the best window size for link prediction.  The table shows that at that window size the other two tasks do not achieve as high a score as if you had chosen the best window size for those two tasks.  This means that the windowing algorithm should choose a different window size for each of the tasks in order to maximize score.

This effect is not limited to just the top-scoring window size.  More generally, the scores between two tasks do not positively correlate with each other, so that a higher score for a given window size on the first task does not necessarily mean a higher score for that window size on the second task.  In order to demonstrate this, we use Spearman's correlation coefficient, which tests the monotonicity between two variables.  A negative score indicates anti-correlation, a score of $0$ indicates no correlation, and a positive score indicates positive correlation.  Table~\ref{table:spearman} show the correlation coefficients and their associated p-values for each pair of tasks:  most of them are either negative or close to zero, indicating little correlation between the tasks.  This is further evidence that the quality of window sizes depend on the task, and hence that we should use the task as supervision for windowing.

\section{Windowing algorithms and baselines}\label{sec:algs}
We now describe the windowing algorithms that we will compare in both the offline and online settings.  So that the supervised algorithms we introduce remain useful for not just the three tasks we consider, but any task, our algorithms do not attempt to take advantage of the particular nature of the task and corresponding algorithm at hand.  In other words, we treat the task algorithms as black boxes.  However, as we show in Section~\ref{sec:results}, the supervised approaches are still able to perform well despite this self-imposed constraint. 
\subsection{Offline supervised algorithms}
Our intuition is very simple:  since we know what task we want to accomplish on the test set, we use the same task algorithm to learn the window size on the training set.  
In the offline setting, in order to try to prevent overfitting and to make the search space smaller, we only consider uniform windowings.  This allows the algorithm to be very simple:  For each window size, up to the length of the training set, window the training set at that size and use that as input for the task algorithm.  Measure the performance of the task algorithm (remember we assume we have ground truth for the training set) and use that as the score for the window size.  Window the test set with the window size that received the highest score.  Of course, this means running the task algorithm $O(T)$ times (where $T$ is the length of the training set), which is not particularly efficient.  However, we make the assumption that since this is an offline setting, this blowup in running time in the training phase is not prohibitively large.  We will refer to this as the offline \emph{supervised} method.

Among the three tasks, the attribute prediction algorithm has an important difference, in so much as that it requires training data to build a model.  We therefore need to make sure to decouple the training data for the model and the training data used to find the best window size.
To do this, we split the training data into two, use the first half as the data for the model, and the second half of the training data to test out how well the model does when windowed at each window size.  We do this by taking the vertices that still have the value of the target attribute and using the same process we use to test the quality of the attribute prediction on the test set:  remove them in batches, build the model at each window size, and then test which value TVRC predicts on the second half of the training data.  As described above, we then use the AUC as the quality of that window size.

\begin{algorithm}
	\caption{Online windowing for link prediction}
	\label{alg:lp_alg}
\begin{algorithmic}
	\REQUIRE Integers $M$ and $B$, Link predictor $L$
	\STATE Initialize $\texttt{scores}_w$ as the empty list
	\FOR{each new graph $G_i$}
		\STATE Let \texttt{new window sizes} include $w$ for $1\le w < i$ if length$(\texttt{scores}_w) < M$
		\STATE Let \texttt{best window sizes} include the top $B$ window sizes by average($\texttt{scores}_w$)
		\FOR{$w$ in \texttt{new window sizes}, \texttt{best window sizes}}
			\STATE Let $\mathcal{H}_w = H_1,\ldots,H_{\lceil \frac{i-1}{w}\rceil}$ be the windowing of $G_1,\ldots,G_{i-1}$ at window size $w$
			\STATE \texttt{predicted links} $= L(\mathcal{H}_w)$
			\STATE Append AUC$(\text{new links in }G_i,\texttt{predicted links})$ to $\texttt{scores}_w$
		\ENDFOR
		\STATE $w^* = \argmax_w \text{average}(\texttt{scores}_w)$
		\STATE Let $\mathcal{H}_{w^*} = H_1,\ldots,H_{\lceil \frac{i}{w^*}\rceil}$ be the windowing of $G_1,\ldots,G_{i}$ at window size $w^*$
		\RETURN $L(\mathcal{H}_{w^*})$
	\ENDFOR
\end{algorithmic}
\end{algorithm}

\begin{table*}
\caption{Performance of each of the algorithms on five data sets with respect to link prediction.}
\centering
\begin{tabular}{cccccc}
\toprule
 & Enron & Reality Mining & Badge & Hypertext	 & Haggle\\
 \midrule
Random & 0.101  & 0.202  & 0.208  & 0.049  & 0.195\\
Hand-picked   & 0.148  & \bf{0.266}  &\bf{0.619}  & \bf{0.146}  &\bf{0.485}\\
\midrule
Weighted Algorithm~\ref{alg:lp_alg} & \bf{0.178}  & \bf{0.263}  & \bf{0.482}  & \bf{0.116}  & 0.443\\
Algorithm \ref{alg:lp_alg} & 0.153  & 0.259  & 0.428  & 0.100  & \bf{0.475}\\
Training only  & 0.159  & 0.240  & 0.418  & 0.065  & 0.437\\
ADAGE  & 0.149  & 0.198  & 0.394  & 0.044  & 0.298\\
\bottomrule
\end{tabular}
\label{table:lp}
\end{table*}

\subsection{Online supervised algorithms}
In the online setting, we could at each time step perform a similar procedure as in the offline case, leading to $O(i)$ runs of the task algorithm at the $i$th step, for a total of $O(T^2)$ times, where $T$ is the total length of the sequence.  This will often be prohibitively expensive.  We introduce an approximate online windowing algorithm to deal with this issue.  We illustrate with link prediction, as a natural example of an online task.

Every time we receive a new graph $G_i$, representing the edges that occurred in the next time step, we can test each window size $w$ by binning the sequence so far at size $w$ and then use the last graph in the windowed sequence to predict the edges that will appear in $G_i$.  We then compare the predicted edges to the actual edges in $G_i$, producing an AUC score for that window size\footnote{For the sake of computational efficiency, we only score pairs of vertices with non-zero degree, since the Katz score will produce a score of 0 for all other pairs.}.  
The window size $w^*$ chosen next to bin the sequence seen so far including $G_i$ is the window size that maximizes the average of all scores for that window size so far.  However, this still means testing $O(i)$ window sizes at each time step, for a total of $O(T^2)$ tests.  

\begin{table}
\caption{Performance of each of the algorithms on three data sets with respect to attribute prediction.}
\centering
\begin{tabular}{cccc}
\toprule
 & Enron & Reality Mining& Badge\\
 \midrule
 Random	& 0.566	& 0.966& 0.583\\
Hand-picked	& 0.560	& 0.960 & 0.646\\
\midrule
Supervised & \textbf{0.587} & \textbf{0.974}& \textbf{0.656}\\
No time	& 0.584	& 0.971 & 0.568\\
Fourier	& 0.567	& 0.967 & 0.568\\
Jaccard	& 0.576	& 0.973 & 0.571\\
Entropy	& 0.555	& 0.970 & 0.562\\
ADAGE	& 0.564	& 0.973 & 0.572\\
\bottomrule
\end{tabular}
\label{table:attr}
\end{table}

To decrease the number of tests, we instead use an approximate version of this where only some of the window sizes are tested, as described in Algorithm~\ref{alg:lp_alg}.  This online algorithm is described explicitly for link prediction, but it is worth nothing that the same algorithm may in principle be used for any online task.  Given hyperparameters to the algorithm $M$ and $B$, we test a window size if it has either been tested fewer than $M$ times, or if the average score so far ranks it amongst the top $B$ window sizes.  The intuition behind this approach is that a window size that has been performing badly will not suddenly become the best performing window size, and thus doesn't need to be tested.  This requires only $O(M\cdot B\cdot T)$ total runs of the link prediction algorithm instead of $O(T^2)$, where we think of as $M$ and $B$ as constants.  In our experiments, we set $M=B=10$ (We also demonstrate the effect of changing these hyperparameters in Section~\ref{sec:results}).  We also test a weighted variant of Algorithm~\ref{alg:lp_alg} (\emph{Weighted Algorithm~\ref{alg:lp_alg}}) by instead using a weighted average of the scores for each window size, in order to privilege scores closer to the present than the past.  We use an exponential weighting scheme, where the weight for the score tested on the $j$th graph where $t$ graphs have been seen already is $\alpha^{t-j}$.  For the purpose of our experiments, we use $\alpha=1/2$.

We also consider a version of this that stops generating new scores after the training period is over, and sticks with the best window size for the training data for all future time steps (\emph{Training only}).

\begin{table}
\caption{Performance of each of the algorithms on two data sets with respect to change point detection.}
\centering
\begin{tabular}{ccc}
\toprule
 & Enron & Reality Mining\\
 \midrule
Random	& 0.660	& 0.609\\
Hand-picked	& 0.700	& 0.643\\
\midrule
Supervised & 0.649	& 0.490\\
Fourier	& 0.657	& 0.405\\
Jaccard	& 0.649	& 0.361\\
Entropy	& 0.709	&\bf{0.891}\\
ADAGE	& \bf{0.766} & 0.356\\
\bottomrule
\end{tabular}
\label{table:cp}
\end{table}

\subsection{Other windowing algorithms and baselines}

We compare against \emph{ADAGE}, the method of Soundarajan et al.~\cite{soundarajan2016generating}.  ADAGE needs a metric as a parameter, so we use the exponent of the degree distribution, which is the metric they use.  We also compare against the Jaccard-index-based method (\emph{Jaccard}) of Darst et al.~\cite{darst2016detection} and the entropy-based method (\emph{Entropy}) of De Domenico et al.~\cite{de2015structural}.  Since this method allows for graphs with any types of layers, we treat each time step as a layer and modify their method to only allow adjacent time steps to be merged.

\begin{figure*}
\centering
\begin{subfigure}{0.3\textwidth}
\includegraphics[width=\columnwidth]{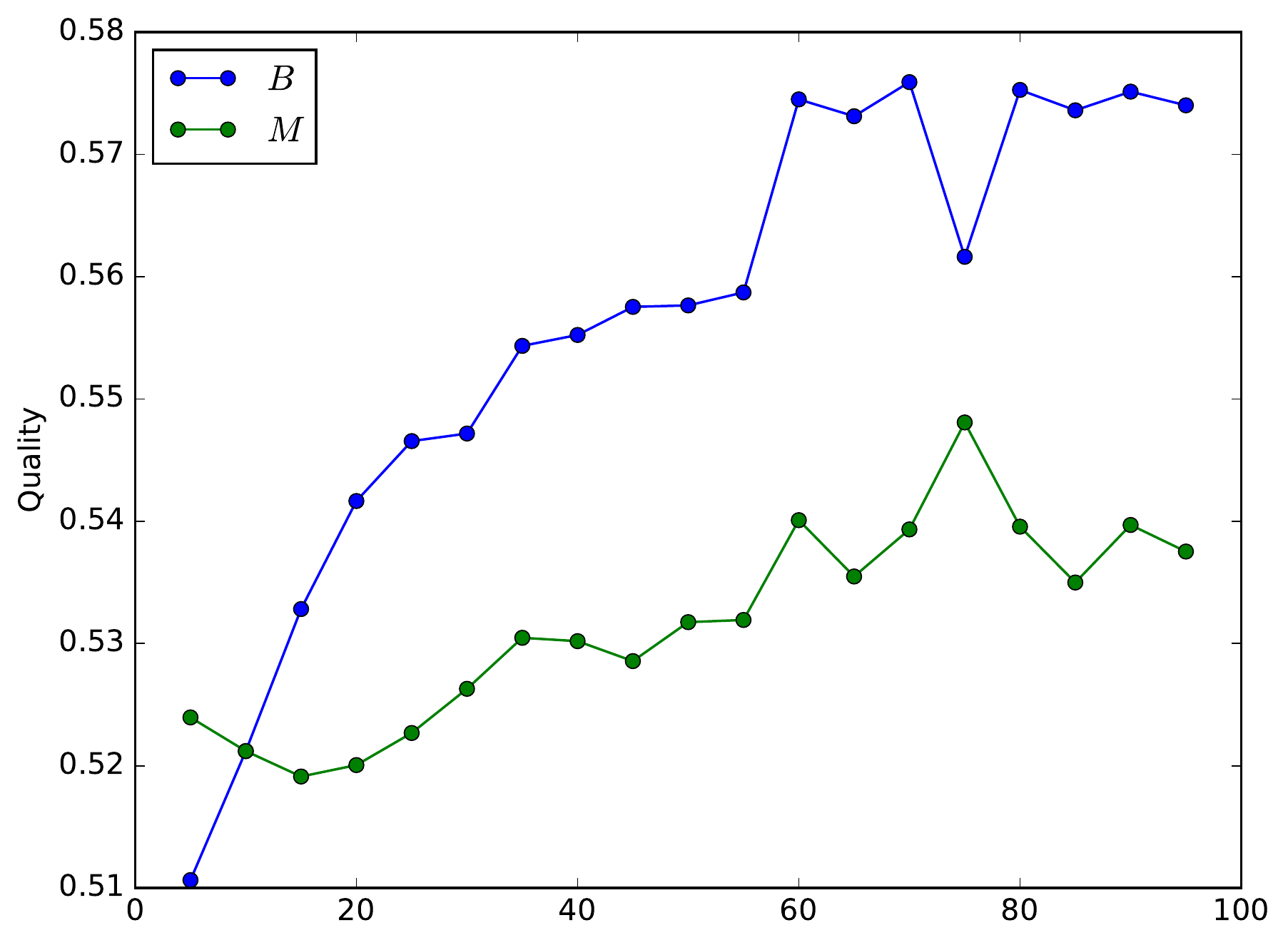}
\end{subfigure}
\begin{subfigure}{0.3\textwidth}
\includegraphics[width=\columnwidth]{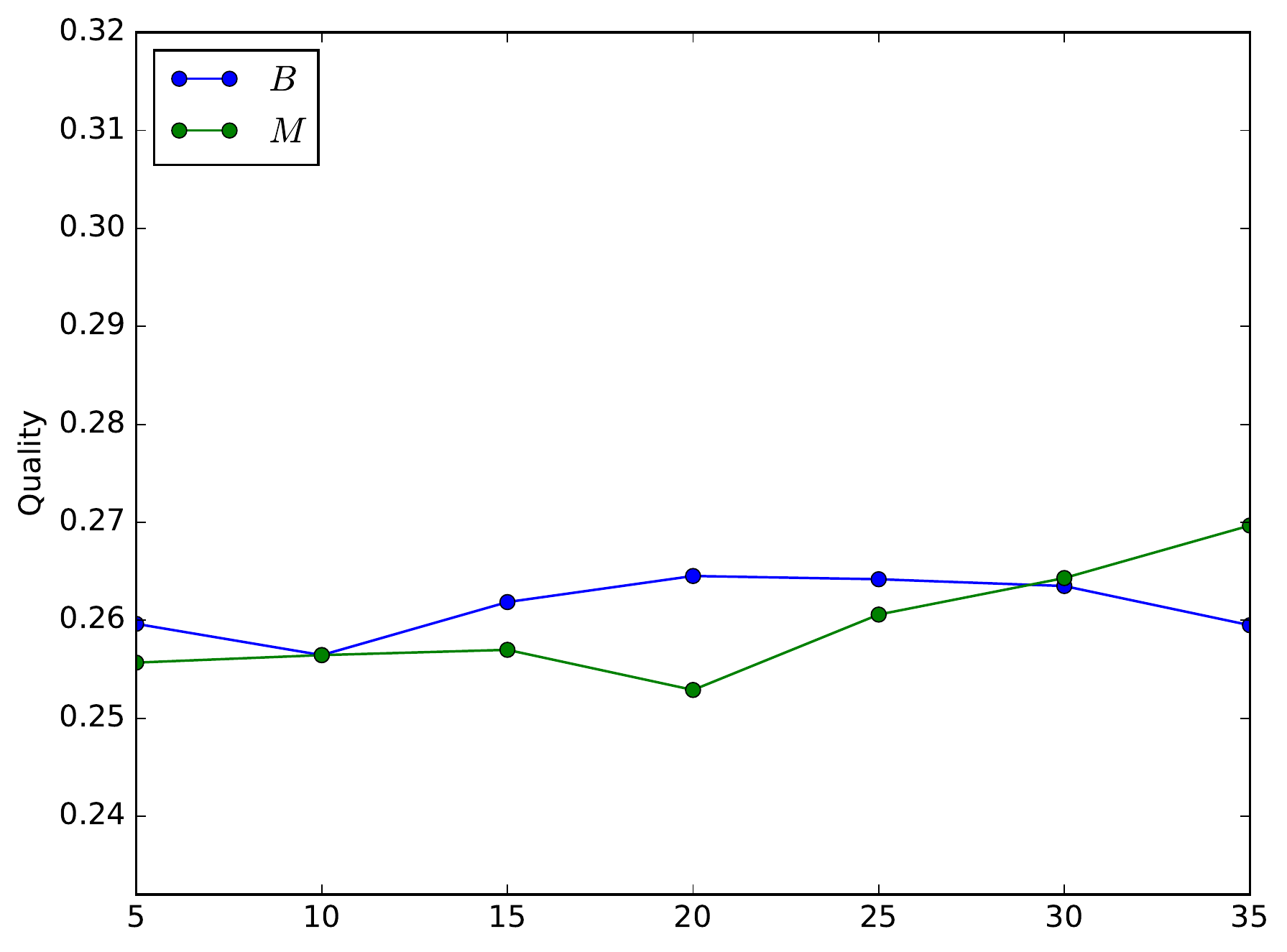}
\end{subfigure}
\caption{The effect of the hyperparameters $B$ and $M$ on quality for link prediction, where we fix $B$=10 and vary $M$, and fix $M=10$ and vary $B$.  The figure on the left is the results on the Badge data set, while the right is the results on the Reality Mining dataset.} 
\label{fig:hyperparams} 
\end{figure*}

We also compare against several baselines:  the first is always using a window size of $w=1$, which as mentioned above, represents a `hand-chosen' value (which we will refer to as the \emph{hand-chosen} algorithm).  This represents the window size that an expert might have chosen.  The second baseline is the random algorithm (\emph{Random}), which chooses a random windowing of the test set.  In addition, for attribute prediction, we also consider the windowing that removes all temporal information, i.e.~the window size that is always the length of the test sequence (\emph{No time}).  The final baseline we consider is slightly more sophisticated:  Consider a time series that assigns a real value to every graph in the sequence.
For any such time series, we may compute its discrete Fourier transform (DFT).  For frequency $f$, denote by $x_f$ the amplitude of that frequency.  The score we assign $w$ is the maximum magnitude $|x_f|$ of any frequency in the transform such that $f$ rounds to $w$.  Only windows where there is such a frequency are assigned scores, and we choose the window size with the maximum score (\emph{Fourier}).  In this paper, we consider the DFT under the commonly-used Hanning window where the metric is the number of edges in each graph.  This serves as a proxy for the amount of activity at any given time.  The DFT of this particular time series for dynamic networks has been used before, e.g.~to study the Reality Mining data set~\cite{eagle2006reality}.

All of these algorithms can be used in the offline setting, but only ADAGE, the hand-chosen baseline, and the random baseline can be used in the online setting.

\begin{figure*}
\centering
\begin{subfigure}{0.32\textwidth}
\includegraphics[width=\columnwidth]{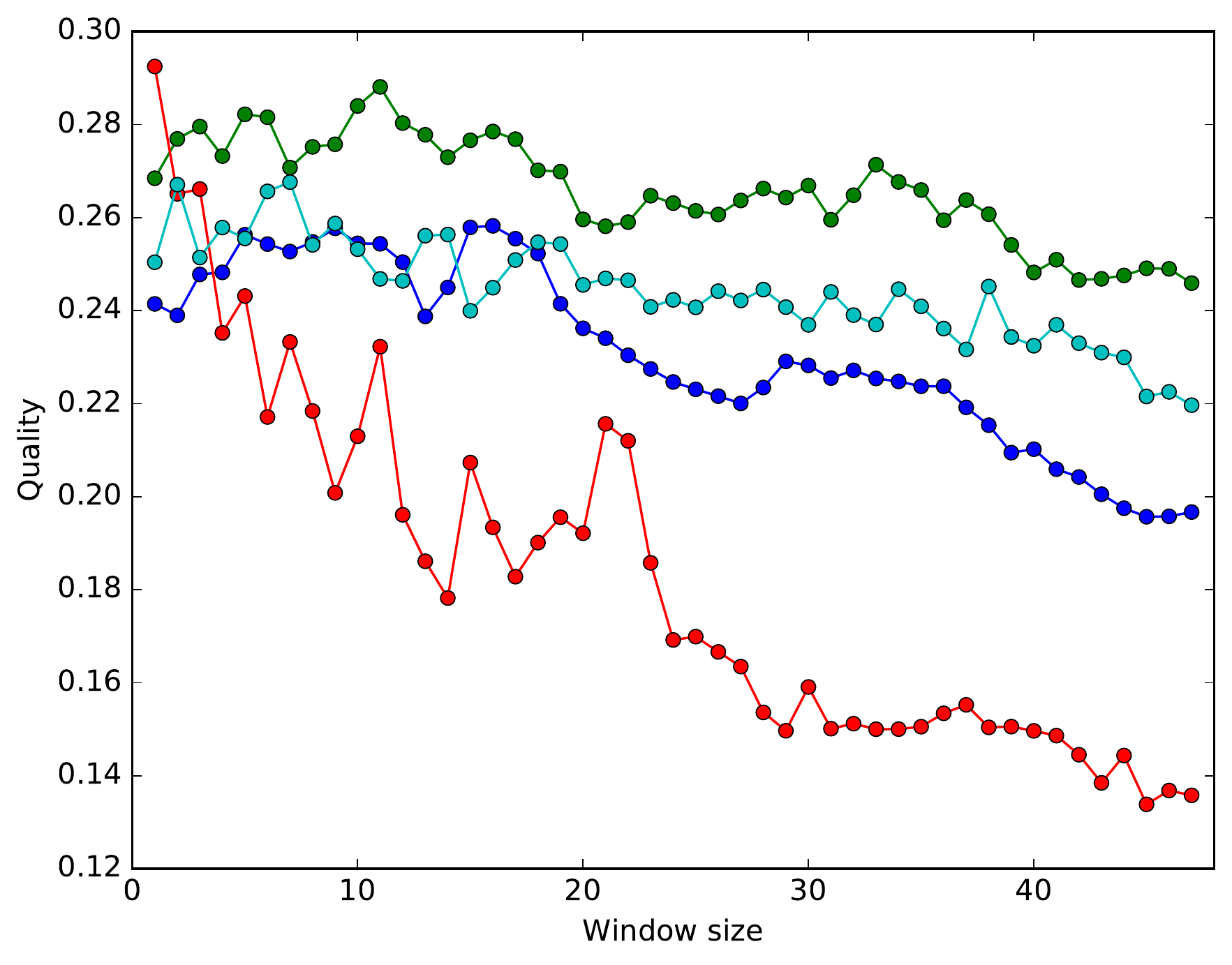}
\end{subfigure}
\hspace{1mm}
\begin{subfigure}{0.32\textwidth}
\includegraphics[width=\columnwidth]{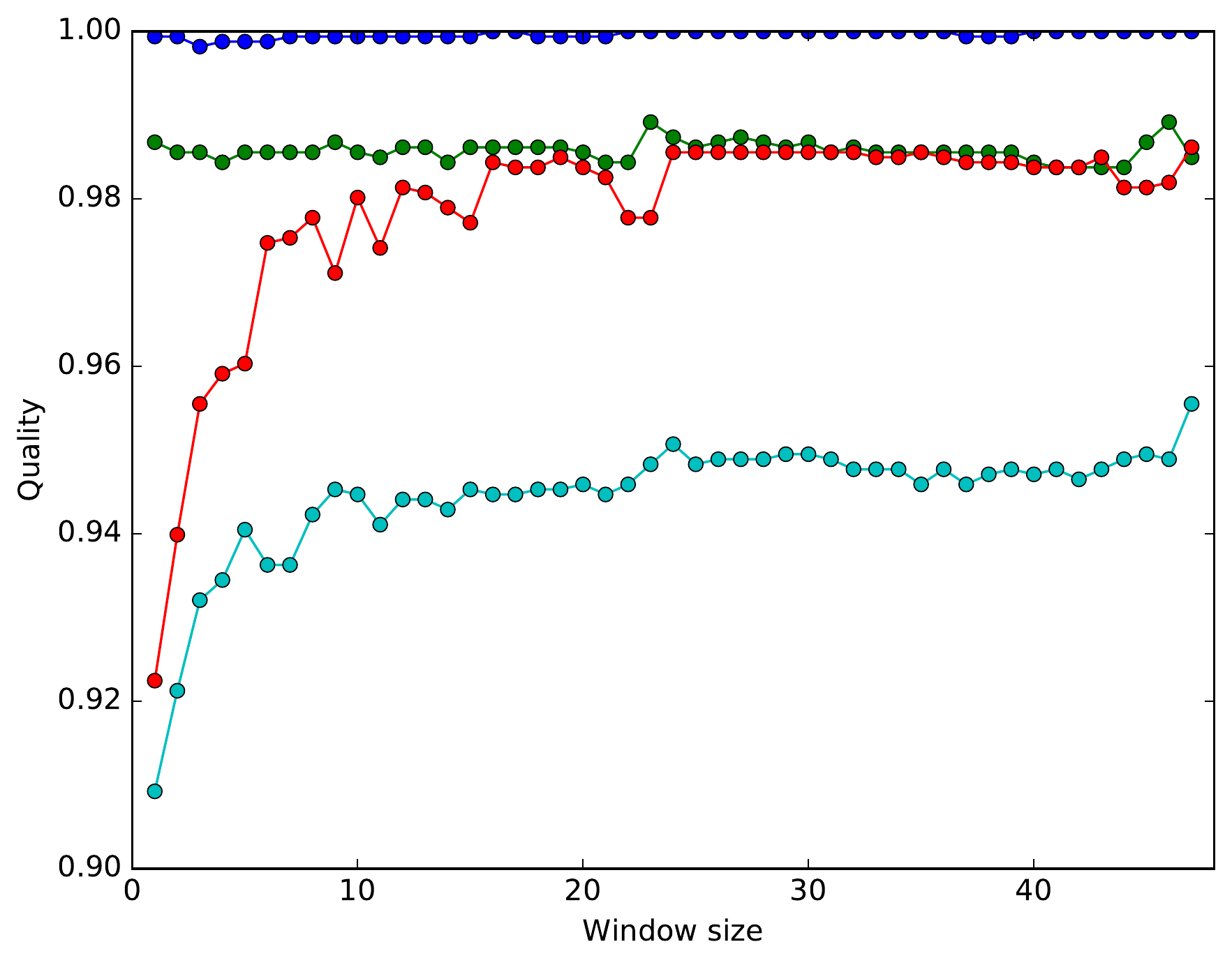}
\end{subfigure}
\hspace{1mm}
\begin{subfigure}{0.32\textwidth}
\includegraphics[width=\columnwidth]{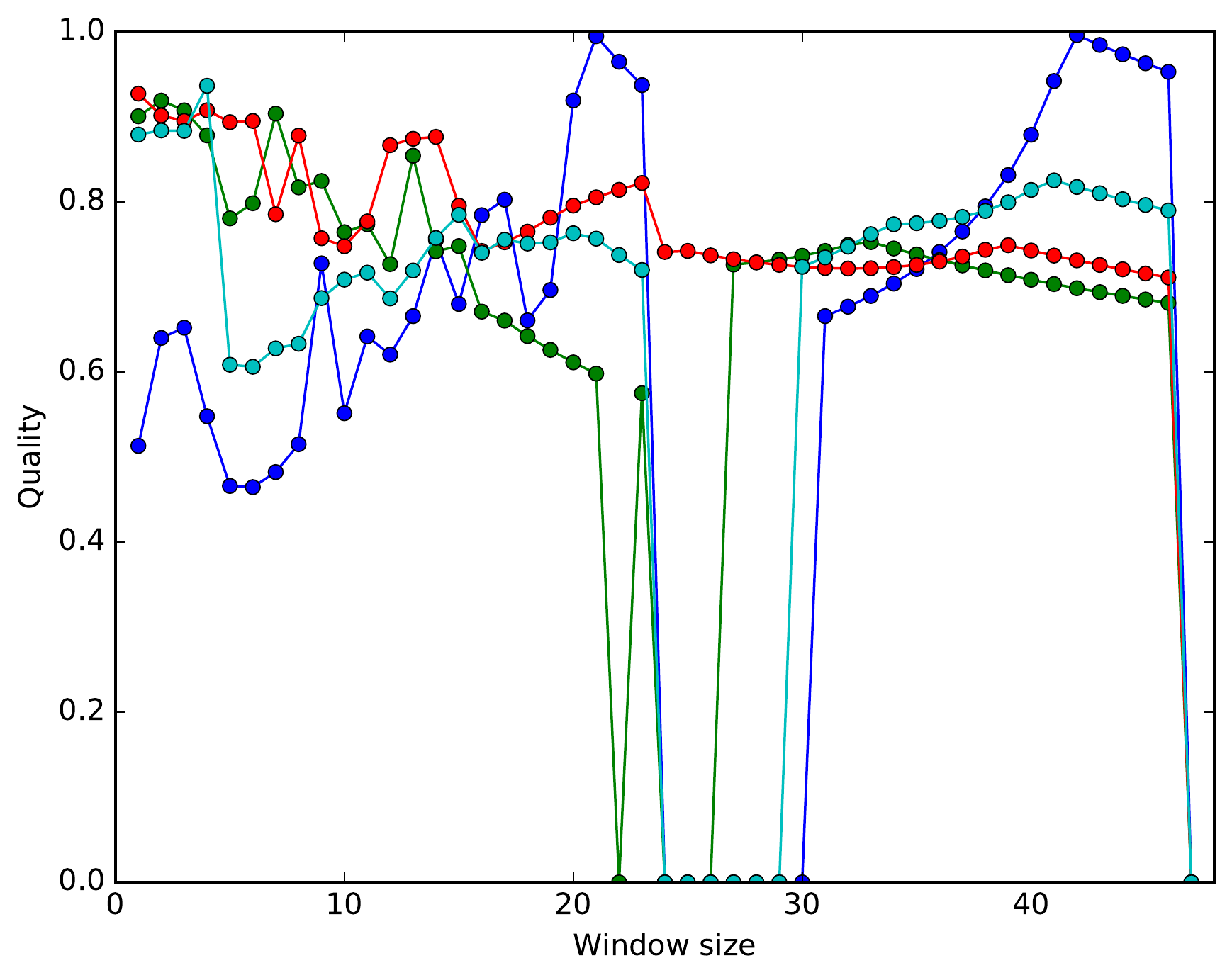}
\end{subfigure}
\caption{Results for each window size on each interval of the Reality Mining data.  Each line corresponds with each of the four intervals.  The left figure shows the quality of each window size for link prediction, the middle for attribute prediction, and the right for change point prediction.} 
\label{fig:rm_uni_splits} 
\end{figure*}

\section{Results}\label{sec:results}

Tables~\ref{table:lp},~\ref{table:attr}, and~\ref{table:cp} show our results.  The supervised approach does do well, but there are certainly caveats.  In link prediction, the `hand-picked' window size is often the best performer, probably due to the fact that these are well-studied data sets with natural periodicities dictated by human activity, and those window sizes reflect that.  However, in general, we will want to have windowing algorithms that do not rely on an analyst's knowledge of the data set:  acquiring domain-specific knowledge like this can be time-consuming, expensive, or difficult.  Outside of this approach, at least one of the supervised methods does the best for all of the data sets.  The weighted version is the overall winner, with a caveat:  like any supervised technique, overfitting is always going to be an issue.  We surmise that this is why the unweighted version outperforms the weighted version on the Haggle data set.

We also test the effect of the hyperparameters $B$ and $M$, shown in Figure~\ref{fig:hyperparams}.  As would be expected, by increasing $B$ or $M$ (which increases the number of window sizes tested) performance is generally improved.  This effect, however, is extremely weak on the Reality Mining data set. That and our ability to perform relatively well even for small values of $B$ and $M$ validates our choice to use small constant values of $B$ and $M$ to use for Algorithm~\ref{alg:lp_alg} and forego any attempt to optimize their values in the course of the windowing algorithm.

In attribute prediction, the supervised approach is the best performing algorithm, although the absolute difference over the others is sometimes rather small.  On the other hand, the supervised approach does not do well for change point prediction.  We believe this is due to a limitation of supervised approaches:  they depend on the quality of the training data.  Given the sparsity of change points, each training set contains only a very small number of change points, making it difficult to distinguish between different window sizes.  This issue separates change point detection from our attribute prediction problem, which has a much greater amount of training data to work with.  

Our approach also reveals other differences between the three tasks.  Figure~\ref{fig:rm_uni_splits} shows the quality of every window size on each of the first four intervals of Reality Mining (we don't use the last interval because each interval needs a subsequent interval for testing attribute prediction).  The scores for change point detection are extremely sensitive to window size, with small differences in size making a large difference, indicating Graphscope's sensitivity to window size.  TVRC, on the other hand, is much more stable under changes to window size, with link prediction falling somewhere in the middle.  Such sensitivity makes it much more difficult to speed up the process by only testing some of the windows sizes instead of all of them, as we do here.  We leave for future work determining if there is a way to test a fewer number of window sizes and if the choice of algorithm for a given task impacts sensitivity to window size.

\begin{figure}
\centering
\includegraphics[width=0.85\columnwidth]{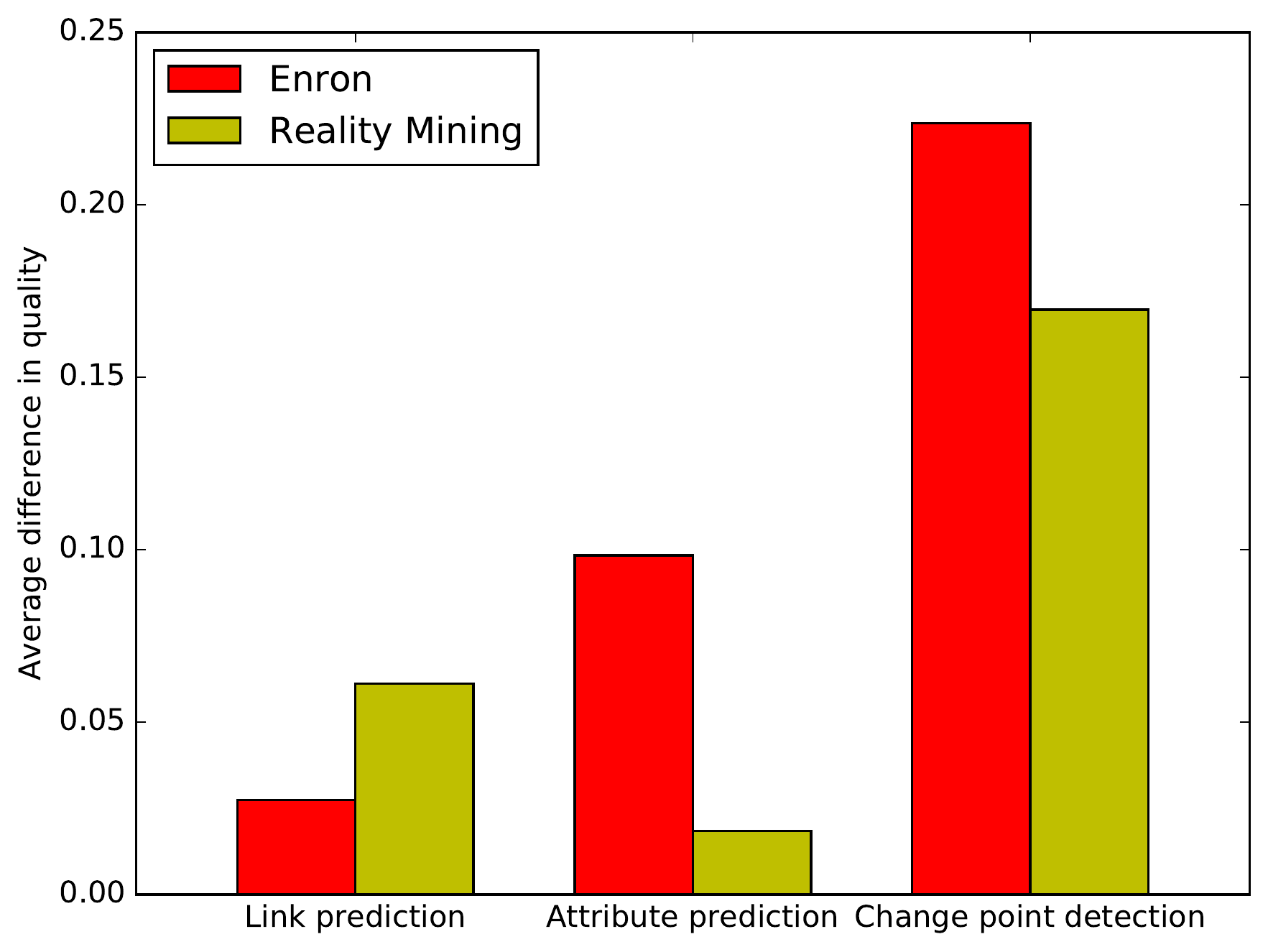}
\caption{For each task, the average difference between quality of the same window size on consecutive intervals of both data sets.}
\label{fig:splits_avg_diff}
\end{figure}

Our supervised approach makes the basic assumption that a window size that works well in the past is more likely to work well in the future.  To measure this, Figure~\ref{fig:splits_avg_diff} shows the difference between the two scores a window size got on consecutive intervals, averaged over all windows and consecutive intervals.  Smaller differences mean that past data reflects future data more precisely.  Figure~\ref{fig:splits_avg_diff} shows the difference was significantly larger for change point detection than the other two tasks, indicating that this assumption (at least when using Graphscope) is violated more under this task.  This serves as evidence for why the supervised approach does not perform as well on change point detection.


%


\section{Conclusion and future work}

In this paper, we have given evidence that windowing is task-dependent.  Recognizing this, we have provided a simple and easy-to-use framework for directly comparing the quality of windowing algorithms and moreover, introduced supervised windowing algorithms that leverage our ability to test a windowing.  Nonetheless, we leave for future work several challenges:  Like any supervised machine learning, the quality of the learner depends on the quantity and quality of training data.  Improving windowing algorithms in the face of environments with little training data remains an issue.  Even with training data, this can be a computationally-expensive procedure if the windowing algorithm has to repeatedly invoke an expensive task algorithm.  We leave for future work finding heuristics that approximate well how a windowing will perform for a given task.  We also leave for future work if performance may be improved by considering other kinds of windowings besides the non-overlapping windowings focused on in this paper.

\bibliographystyle{plain}
\bibliography{cits}

\begin{thebibliography}{10}

\bibitem{al2011survey}
Mohammad Al~Hasan and Mohammed~J. Zaki.
\newblock A survey of link prediction in social networks.
\newblock In {\em Social Network Data Analytics}, pages 243--275. Springer,
  2011.

\bibitem{berlingerio2010time}
Michele Berlingerio, Michele Coscia, Fosca Giannotti, Anna Monreale, and Dino
  Pedreschi.
\newblock As time goes by: Discovering eras in evolving social networks.
\newblock In {\em Pacific-Asia Conference on Knowledge Discovery and Data
  Mining}, pages 81--90. Springer, 2010.

\bibitem{clauset2012persistence}
Aaron Clauset and Nathan Eagle.
\newblock Persistence and periodicity in a dynamic proximity network.
\newblock {\em {DIMACS} Workshop on Computational Methods for Dynamic
  Interaction Networks}, 2007.

\bibitem{creamer2009segmentation}
Germ{\'a}n Creamer, Ryan Rowe, Shlomo Hershkop, and Salvatore~J Stolfo.
\newblock Segmentation and automated social hierarchy detection through email
  network analysis.
\newblock In {\em Advances in Web Mining and Web Usage Analysis}, pages 40--58.
  Springer, 2009.

\bibitem{darst2016detection}
Richard~K. Darst, Clara Granell, Alex Arenas, Sergio G{\'o}mez, Jari
  Saram{\"a}ki, and Santo Fortunato.
\newblock Detection of timescales in evolving complex systems.
\newblock {\em arXiv preprint arXiv:1604.00758}, 2016.

\bibitem{de2015structural}
Manlio De~Domenico, Vincenzo Nicosia, Alexandre Arenas, and Vito Latora.
\newblock Structural reducibility of multilayer networks.
\newblock {\em Nature Communications}, 6, 2015.

\bibitem{eagle2006reality}
Nathan Eagle and Alex~Sandy Pentland.
\newblock Reality mining: sensing complex social systems.
\newblock {\em Personal and Ubiquitous Computing}, 10(4):255--268, 2006.

\bibitem{fish2015handling}
Benjamin Fish and Rajmonda~S. Caceres.
\newblock Handling oversampling in dynamic networks using link prediction.
\newblock In {\em Joint European Conference on Machine Learning and Knowledge
  Discovery in Databases}, pages 671--686. Springer, 2015.

\bibitem{isella2011hypertext}
Lorenzo Isella, Juliette Stehl{\'e}, Alain Barrat, Ciro Cattuto,
  Jean-Fran{\c{c}}ois Pinton, and Wouter Van~den Broeck.
\newblock What's in a crowd? analysis of face-to-face behavioral networks.
\newblock {\em Journal of Theoretical Biology}, 271(1):166--180, 2011.

\bibitem{keogh2004segmenting}
Eamonn Keogh, Selina Chu, David Hart, and Michael Pazzani.
\newblock Segmenting time series: A survey and novel approach.
\newblock {\em Data mining in Time Series Databases}, 57:1--22, 2004.

\bibitem{klimt2004enron}
Bryan Klimt and Yiming Yang.
\newblock The enron corpus: A new dataset for email classification research.
\newblock In {\em European Conference on Machine Learning}, pages 217--226.
  Springer, 2004.

\bibitem{krings2012effects}
Gautier Krings, M{\'a}rton Karsai, Sebastian Bernhardsson, Vincent~D Blondel,
  and Jari Saram{\"a}ki.
\newblock Effects of time window size and placement on the structure of an
  aggregated communication network.
\newblock {\em EPJ Data Science}, 1(1):4, 2012.

\bibitem{LibenNowell2003}
David Liben-Nowell and Jon Kleinberg.
\newblock The link prediction problem for social networks.
\newblock In {\em Proceedings of the Twelfth International Conference on
  Information and Knowledge Management}, CIKM '03, pages 556--559, New York,
  NY, USA, 2003. ACM.

\bibitem{liu2016graph}
Yike Liu, Abhilash Dighe, Tara Safavi, and Danai Koutra.
\newblock A graph summarization: A survey.
\newblock {\em arXiv preprint arXiv:1612.04883}, 2016.

\bibitem{olguin2009badge}
Daniel~Olgu{\'\i}n Olgu{\'\i}n, Benjamin~N. Waber, Taemie Kim, Akshay Mohan,
  Koji Ara, and Alex Pentland.
\newblock Sensible organizations: Technology and methodology for automatically
  measuring organizational behavior.
\newblock {\em IEEE Transactions on Systems, Man, and Cybernetics, Part B
  (Cybernetics)}, 39(1):43--55, 2009.

\bibitem{peel2015detecting}
Leto Peel and Aaron Clauset.
\newblock Detecting change points in the large-scale structure of evolving
  networks.
\newblock In {\em Twenty-Ninth AAAI Conference on Artificial Intelligence},
  2015.

\bibitem{ribeiro2013quantifying}
Bruno Ribeiro, Nicola Perra, and Andrea Baronchelli.
\newblock Quantifying the effect of temporal resolution on time-varying
  networks.
\newblock {\em Scientific Reports}, 3:3006, 2013.

\bibitem{haggle}
James Scott, Richard Gass, Jon Crowcroft, Pan Hui, Christophe Diot, and
  Augustin Chaintreau.
\newblock {CRAWDAD} dataset cambridge/haggle (v. 2009-05-29).
\newblock Downloaded from \url{http://crawdad.org/cambridge/haggle/20090529},
  May 2009.

\bibitem{sharan2008temporal}
Umang Sharan and Jennifer Neville.
\newblock Temporal-relational classifiers for prediction in evolving domains.
\newblock In {\em Eighth IEEE International Conference on Data Mining, 2008.
  ICDM'08}, pages 540--549. IEEE, 2008.

\bibitem{soundarajan2016generating}
Sucheta Soundarajan, Acar Tamersoy, Elias~B. Khalil, Tina Eliassi-Rad,
  Duen~Horng Chau, Brian Gallagher, and Kevin Roundy.
\newblock Generating graph snapshots from streaming edge data.
\newblock In {\em Proceedings of the 25th International Conference Companion on
  World Wide Web}, pages 109--110. International World Wide Web Conferences
  Steering Committee, 2016.

\bibitem{sulo2010meaningful}
Rajmonda Sulo, Tanya Berger-Wolf, and Robert Grossman.
\newblock Meaningful selection of temporal resolution for dynamic networks.
\newblock In {\em Proceedings of the Eighth Workshop on Mining and Learning
  with Graphs}, pages 127--136. ACM, 2010.

\bibitem{sun2007graphscope}
Jimeng Sun, Christos Faloutsos, Spiros Papadimitriou, and Philip~S. Yu.
\newblock Graphscope: parameter-free mining of large time-evolving graphs.
\newblock In {\em Proceedings of the 13th ACM SIGKDD International Conference
  on Knowledge Discovery and Data Mining}, pages 687--696. ACM, 2007.

\bibitem{yang2015evaluating}
Yang Yang, Ryan~N. Lichtenwalter, and Nitesh~V. Chawla.
\newblock Evaluating link prediction methods.
\newblock {\em Knowledge and Information Systems}, 45(3):751--782, 2015.

\end{thebibliography}

\end{document}